\documentclass[aps,pre,superscriptaddress,unsortedaddress,twocolumn]{revtex4}

\usepackage{amsfonts}

\usepackage{graphicx}
\usepackage{epstopdf}
\usepackage{amsmath}
\usepackage{amssymb}
\usepackage{verbatim}
\usepackage[ulem=normalem]{changes}
\usepackage{booktabs}
\usepackage{natbib}

\begin{document}

\title{Measuring kinetic energy changes in the mesoscale\\
with low acquisition rates\cite{memory}}
\date{\today}

\small

\begin{abstract}

We report on the first measurement of the average kinetic energy changes in isothermal and non-isothermal quasistatic processes in the mesoscale, realized with a Brownian particle trapped with optical tweezers. Our estimation of the kinetic energy change allows to access to the full energetic description of the Brownian particle. Kinetic energy estimates are obtained from measurements of the mean square velocity of the trapped bead sampled at frequencies several orders of magnitude smaller than the momentum relaxation frequency. The velocity is tuned applying a noisy electric field that modulates the amplitude of the fluctuations of the position and velocity of the Brownian particle, whose motion is equivalent to that of a particle in a higher temperature reservoir.  Additionally, we show that the dependence of the variance of the time-averaged velocity on the sampling frequency can be used to quantify properties of the electrophoretic mobility of a charged colloid. Our method could be applied to detect temperature gradients in inhomogeneous media and to characterize the complete thermodynamics of biological motors and of artificial micro and nanoscopic heat engines.
 
 \end{abstract}


\author{\'E. Rold\'an$^{1,2}$, I. A. Mart\'inez$^1$, L. Dinis$^{2,3}$ and R. A. Rica$^1$\email{rul@ugr.es}}
\affiliation{$^1$ICFO $-$ Institut de Ci\`encies Fot\`oniques, Mediterranean Technology Park,  Av. Carl Friedrich Gauss, 3, 08860, Castelldefels (Barcelona), Spain. \\
$^2$GISC $-$ Grupo Interdisciplinar de Sistemas Complejos. Madrid, Spain.\\
$^3$Departamento de F\'isica At\'omica, Molecular y Nuclear, Universidad Complutense de Madrid, 28040,  Madrid, Spain.}

\maketitle





Colloidal particles suspended in fluids are subject to thermal fluctuations that produce a random motion of the particle, which was firstly observed by Brown~\cite{Brown1828} and described by Einstein's theory~\cite{EinsteinAnnPhys1905}. Fast impacts from the molecules of the surrounding liquid induce an erratic motion of the particle, which returns momentum to the fluid at times $\sim m/\gamma$, $m$ being the mass of the particle and $\gamma$ the friction coefficient~\cite{Langevin1908}, thus defining an inertial characteristic frequency $f_p=\gamma/2\pi m$. The momentum relaxation time is of the order of nanoseconds for the case of microscopic dielectric beads immersed in water. Therefore, in order to accurately measure the instantaneous velocity of a Brownian particle, it is necessary to sample the position of the particle with sub-nanometer precision and at a sampling rate above MHz. 

Experimental access of the instantaneous velocity of Brownian particles is of paramount importance not only for the understanding of Brownian motion. The variations of kinetic energy that occur at the mesoscale are relevant for the detailed description of Stochastic energetics~\cite{Sekimoto2010, Schmiedl2008}, the notion of entropy at small scales~\cite{seifert2005entropy,Dunkel2005time} and the statement of fluctuation theorems~\cite{seifert2012stochastic}. A correct energetic description of micro and nano heat engines would only be possible taking into account the kinetic energy changes~\cite{sekimoto2000carnot,Schmiedl2008,Blickle2011}. Applications on the microrheology of complex media~\cite{Raj2013Studying}, the study of hydrodynamic interactions between suspended colloids~\cite{mittal2008} and single colloid electrophoresis~\cite{Semenov2009Single,Heinengen2010Dynamics,Strubbe2013Electrophoretic,Pesce2013Optical} would also benefit from the experimental access of Brownian motion at short time scales.

Optical tweezers constitute an excellent and versatile tool to manipulate and study the Brownian motion of colloidal particles individually~\cite{AshkinOL1986,CilibertoJSM2010}. State of the art technical capabilities recently allowed to explore time scales at which the motion of Brownian particles is ballistic~\cite{li2010measurement,Huang2011Direct,Kheifets2014Measurement}. It would be interesting however to find a technique that allows one to estimate the velocity of Brownian particles from data acquired at slower rates than MHz, i.e., with a less demanding technology.

\begin{figure}[h!]
\includegraphics[width=8cm]{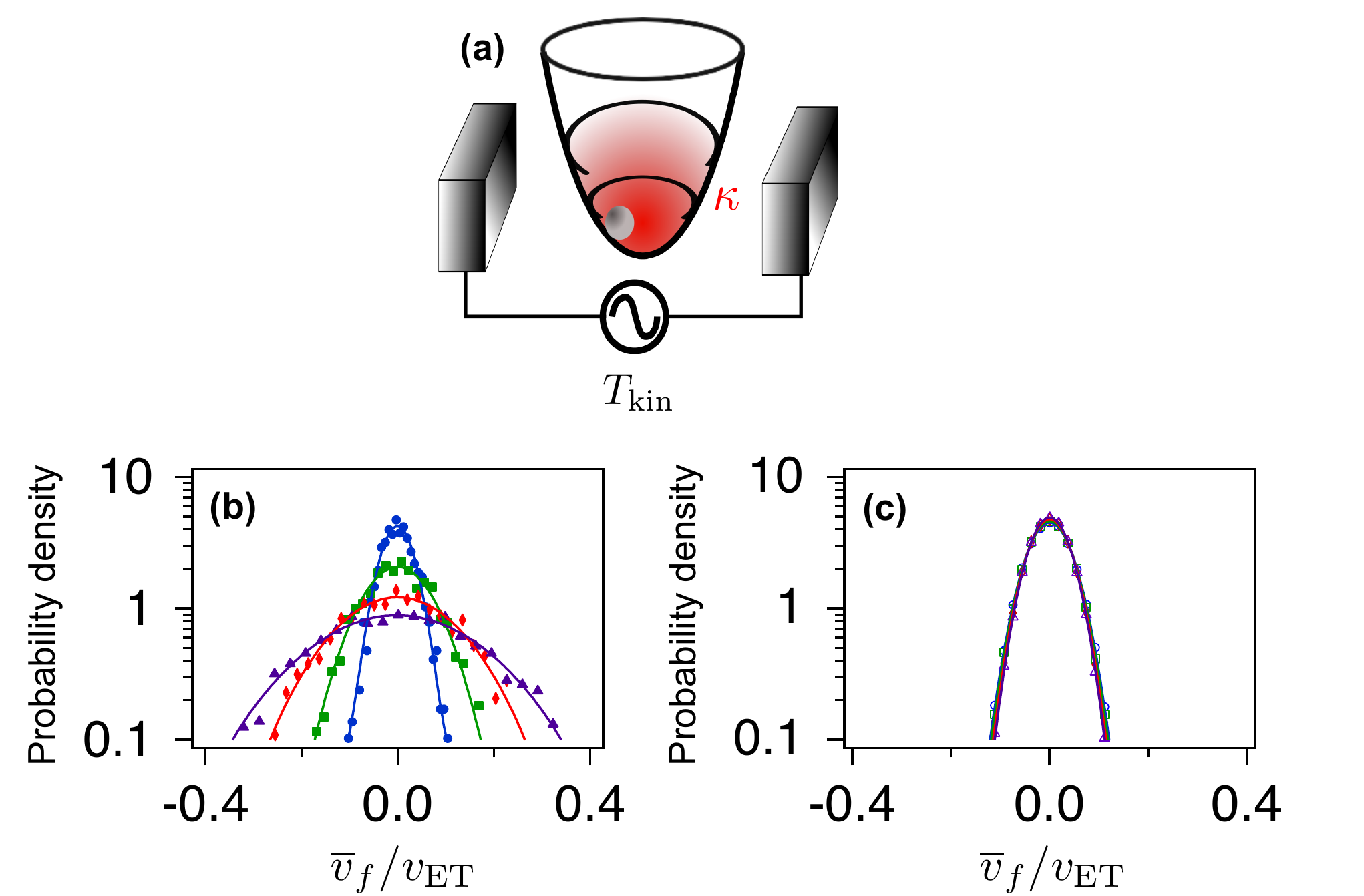}
\caption{{\bf (a)} Schematic picture of our experiment. The stiffness of the trap where a single bead is optically confined is controlled through the power of the trapping laser. The kinetic temperature of the bead is increased by applying a noisy electric field. {\bf (b)} Probability density function of the  time averaged velocity (TAV) normalized to the standard deviation predicted from equipartition $v_{\rm ET} = \sqrt {kT/m}$. Different curves are obtained for the same stiffness $\kappa=(18.0 \pm 0.2)\,\rm pN/\mu m$ but for increasing values of the amplitude of the electric noise, corresponding to the following kinetic temperatures: $T_{\rm kin}=300\, \rm K$ (blue circles), $1060\, \rm K$ (green squares), $3100\, \rm K$ (red diamonds), $5680\, \rm K$ (magenta triangles). Each curve is obtained from a single measurement of $\tau=12\rm s$, sampled at $f = 5\, \rm kHz$. Solid lines are fits to Gaussian distributions. {\bf (c)} The same as {\bf (b)}, but for a fixed kinetic temperature $(T_{\rm kin}=T)$ and three different values of the stiffness: $\kappa=(5.0\pm 0.2) \, \rm pN/\mu m$ (blue circles), $\kappa=(18.7\pm 0.2) \, \rm pN/\mu m$ (green squares), $\kappa=(25.0\pm 0.2) \, \rm pN/\mu m$ (red diamonds) and $\kappa=(38.7\pm 0.2)  \, \rm pN/\mu m$ (magenta triangles).}
\label{fig:overview}
\end{figure}

In this Letter, we estimate the kinetic energy of an optically-trapped colloidal particle immersed in water at different values of its kinetic temperature. Such kinetic temperature is controlled by means of an external random electric field, as described in~\cite{Martinez2013}. The mean squared \emph{instantaneous} velocity of the Brownian particle $-$ or equivalently of its kinetic energy $-$ is estimated from measurements of the average velocity of the particle at sampling rates far below the momentum relaxation frequency. We report measurements performed in equilibrium as well as along isothermal and non-isothermal quasistatic processes. A careful analysis of the experimental results also provides information of the dynamic electrophoretic response of the particle, showing evidences of a low frequency relaxation process.

Figure~\ref{fig:overview} shows a depiction of our experiment, which has been previously described~\cite{Tonin2010,Roldan2013} and is presented in more detail in the Supplementary Material~\cite{supplemental}. A single polystyrene sphere of radius $R=500\,\rm nm$ is immersed in water and trapped with an optical tweezer created by an infrared diode laser. 
A couple of aluminum electrodes located at the ends of a custom-made fluid chamber are used to apply a voltage of controllable amplitude. The key point of our experiment is the simultaneous and accurate control of the two parameters that determine the energy exchanges between a trapped Brownian particle and its environment. Firstly, the optical potential created by the trap is quadratic around its center, $U(x)=\frac{1}{2}\kappa x^2$, $x$ being the position of the particle with respect to the trap centre and $\kappa$ the stiffness of the trap. The trap stiffness can be modulated by tuning the intensity of the trapping laser. Secondly, the kinetic temperature of the particle can be controlled with the amplitude of an external random electric field applied to the electrodes. In the absence of the field, the fluctuations of the position of the particle obey equipartition theorem~\cite{greiner1999thermodynamics},  $\kappa \langle x^2 \rangle = k T$, $k$ being Boltzmann's constant and $T$ the temperature of the water. By applying a random force characterized by a Gaussian white noise process of amplitude $\sigma$, we can mimic the kicks of the solvent molecules to the bead in a higher temperature reservoir, defining a kinetic temperature from the position fluctuations
\begin{equation}
T_{\rm kin} = \frac{\kappa \langle x^2\rangle}{k}.
\label{eq:Tkin}
\end{equation} 
In this situation, $T_{\rm kin} = T + \sigma^2 / 2\gamma k>T$~\cite{Martinez2013,supplemental}. In our experiment, we have control over $\sigma$ through its linear dependence on the voltage applied to the electrodes, and thus we can tune the kinetic temperature of the bead through the amplitude of the external field. Note that we checked that resistive heating is negligible in our experiments taking advantage of the asymmetry of the kinetic temperature, since it is only increased in the direction along which the field is applied. In fact, the continuous monitoring of the temperature in the direction perpendicular to the field did not show any significant deviation from the expected value $T_{\rm kin}=T$~\cite{Martinez2013}.

Using our setup, we can realize any thermodynamic process in which the stiffness of the trap and the kinetic temperature of the particle can change with time arbitrarily following a protocol $\{\kappa(t),T_{\rm kin} (t)\}$. The change in the potential  energy of the particle in the time interval $[t, t+dt]$ is $dU (t) = \left(x^2(t)/2\right) d\kappa (t) + \left(\kappa(t)/2\right)d\left[x^2(t)\right]$. Stochastic energetics defines a framework where these energy changes can be interpreted in terms of work and heat in the meso-scale~\cite{Sekimoto2010}. Therefore, the first term in that equality is the energy change due to the modification of the external controllable parameter (stiffness of the trap), which can be interpreted as the work done on the particle $d'W = \frac{\partial U(x,t)}{\partial \kappa(t)} d\kappa(t)$. The second term is therefore the heat absorbed by the particle  $d'Q = \left(\kappa(t)/2\right)d\left[x^2(t)\right]$~\cite{Sekimoto2010}.

The ensemble average of any thermodynamic quantity ($\langle . \rangle$) is equal to the trajectory-dependent magnitude averaged over all the observed trajectories. The work and heat along a quasistatic process of total duration $\tau$ averaged over many realizations are equal to
\begin{equation}
\langle W \rangle = \int_{0}^{\tau} \frac{ \langle x^2(t)\rangle}{2} d\kappa(t)  = 
\int_{0}^{\tau}   \frac{kT_{\rm kin}(t)}{2\kappa(t)}\,d\kappa(t),
\label{work}
\end{equation}
\begin{equation}
\langle Q \rangle = \int_{0}^{\tau} \frac{\kappa(t)}{2}\, d\left[\langle x^2(t)\rangle\right]  = 
\int_{0}^{\tau}  \frac{\kappa(t)}{2}\, d\left(\frac{kT_{\rm kin}(t)}{\kappa(t)}\right),
\label{heatX}
\end{equation}
where we have used  equipartition theorem along the quasistatic protocols, $\kappa(t)\langle x^2 (t)\rangle  = k T_{\rm kin} (t)$.  The average potential and kinetic energy changes are equal to  $\langle \Delta U \rangle= \langle \Delta E_{\rm kin} \rangle=  \frac{k}{2} [T_{\rm kin}(\tau)-T_{\rm kin}(0)]$. Finally, the total energy change is $\langle \Delta E_{\rm tot} \rangle = \langle \Delta U \rangle + \langle \Delta E_{\rm kin} \rangle=  k [T_{\rm kin}(\tau)-T_{\rm kin}(0)]$. Therefore, for any protocol where $\kappa$ and $T_{\rm kin}$ are changed in a controlled way, all the values of the energy exchanges are known, and can therefore be compared with the measurements we present below. 

Stochastic energetics~\cite{Sekimoto2010,Roldan2013} provides the appropriate framework to measure the infinitesimal exchanges of work $d'W$ and heat $d'Q$ done on the bead from  $t$ to $t+\Delta t$. The complete energetic study requires also the measurement of the instantaneous velocity and of the kinetic energy exchanges. In fact, the contribution of kinetic energy cannot be neglected in any thermodynamic process that involves temporal~\cite{Schmiedl2008} or spatial~\cite{Bo2013} temperature changes. In our experiment, we do not have direct access to the instantaneous velocity due to our limited sampling speed. We have developed a technique that allows to extrapolate the instantaneous velocity from the time averaged velocity (TAV) $\overline{v}_f$ over a time $\Delta t=1/f$. The energy exchanges $d'W$ and $d'Q$, and the TAV are obtained from the measured time series of $n$ data $\{x(i\Delta t)\}_{i=0}^n$, as described in the Supplementary Material~\cite{supplemental}.

\begin{figure}
\includegraphics[height=5cm]{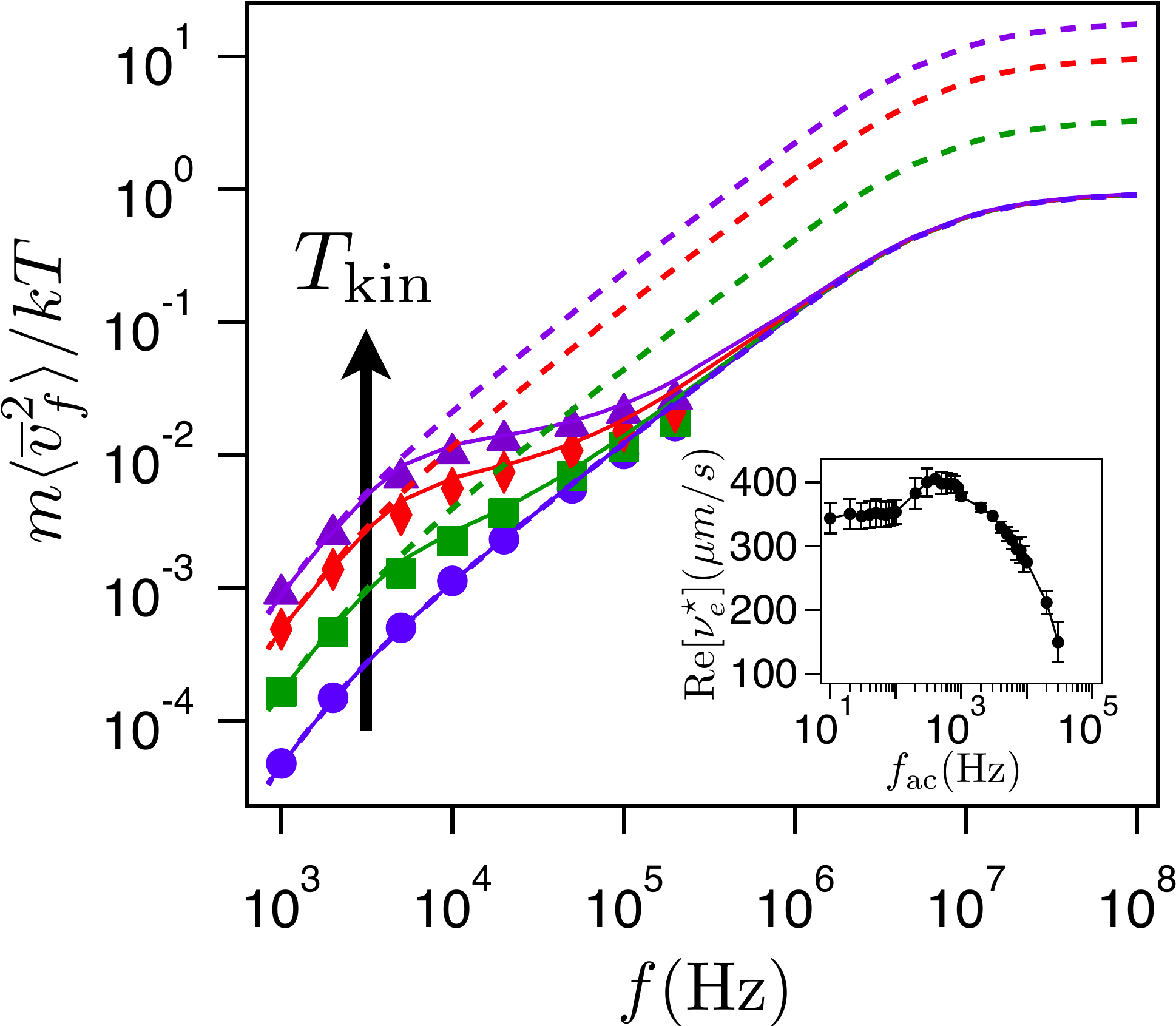} 
\caption{Average kinetic energy measured from the TAV, $\frac{1}{2}m\langle \overline{v}_f^2\rangle$ in units of $\frac{1}{2}(kT)$ as a function of the sampling rate. Different symbols indicate values obtained from the same experiments as those shown in Fig.~\ref{fig:overview}(b). The arrow points towards larger values of $T_{\rm kin}$. The expected theoretical values of $m\langle \overline{v}_f^2\rangle / kT$ for the case of an external field described by a Gaussian white noise are plotted with dashed curves.  Solid curves are obtained numerically for the case of an external noise with flat spectrum up to a  cutoff frequency of $f_{\rm c} = 3\,\rm kHz$. Inset: Real part of the electrophoretic velocity as a function of the frequency $f_{ac}$ of a sinusoidal electric signal of amplitude $200\,\rm V$. The solid line is a guide to the eye. }
\label{fig:Tv_vs_tau}
\end{figure}

The starting point of our study is the analysis of the histograms of the TAV for different values of $\kappa$ and $T_{\rm kin}$. As shown in Eq.~\eqref{eq:Tkin}, the values of $T_{\rm kin}$ are obtained from the variance of the position of the particle. Figure~\ref{fig:overview}(c) shows that the histogram of the TAV is not modified if $\kappa$ is changed keeping $T_{\rm kin}$ constant, while a significant variation of the histogram, and in particular of the variance $\langle \overline{v}^2_{f}\rangle$, is observed if $\kappa$ is fixed and $T_{\rm kin}$ changes from $300\, \rm K$ to $6000\, \rm K$, c.f. Fig.~\ref{fig:overview}(b). Due to the diffusive nature of Brownian motion, the amplitude of the velocities measured at sampling rates below the momentum relaxation frequency $f_p$ are several orders of magnitude below to what is predicted by equipartition theorem, $v_{\rm ET} = \sqrt {kT/m}$~\cite{kerker1974brownian,li2010measurement}.
However, the fact that the variance of the histograms in Fig.~\ref{fig:overview}(b) changes with the noise intensity suggests that there is some information about $T_{\rm kin}$ that can be obtained from them. Notice that the average instantaneous velocity satisfies the equipartition theorem at $T_{\rm kin}$ in our setup, $m\langle v^2\rangle = kT_{\rm kin}$, its variance being independent on $\kappa$.

We now investigate the dependence of the variance of the TAV on the data acquisition frequency, $f$. In Fig.~\ref{fig:Tv_vs_tau} we plot the values of $m\langle \overline{v}^2_f \rangle / kT$ for different values of the noise intensity as functions of $f$. The results for different sampling rates were obtained from the very same time series but sampled at different frequencies ranging from $1\,\rm kHz$ to $200 \,\rm kHz$. For every value of $f$, $\langle \overline{v}_f^2\rangle$ increases with the noise intensity, but all the curves collapse to the same value for high acquisition rates. The underdamped Langevin equation for a Brownian particle trapped in a harmonic potential can help to understand this behavior. 
In the Supplementary Material~\cite{supplemental}, we prove that the variance of the TAV satisfies a modified equipartition theorem,
\begin{equation}
\frac{m\langle \overline{v}^2_{f}\rangle}{kT_{\rm kin}}= L(f),
\label{eq:MET}
\end{equation}
where
\begin{equation}
L(f)= 2f^2\left[ \frac{1}{f_0^2} + \frac{e^{-\frac{f_p}{2f}}}{f_1} \left(  \frac{e^{-f_1/f}}{f_p+2f_1} - \frac{e^{f_1/f}}{f_p-2f_1}  \right)\right],
\label{eq:FactorBdeC_trapped}
\end{equation}
$f_0=\sqrt{f_p f_\kappa}$, $f_\kappa=\kappa/2\pi\gamma$ and $f_1=\sqrt{f_p^2/4-f_0^2}$. The function $L(f)$ is thus a measure of the deviation from the equipartition theorem. It satisfies $L(f)<1$ for any sampling frequency below $f_p$ and its asymptotic behavior is in accordance with equipartition theorem, i.e., $L(f)\to 1$ when $f\to \infty$.

Equations~\eqref{eq:MET} and \eqref{eq:FactorBdeC_trapped} reproduce the observed experimental data without any fitting parameter when $f$ is sufficiently low, as shown by the dashed curves in Fig.~\ref{fig:Tv_vs_tau}. However, the measured value of $\langle \overline{v}_f^2\rangle$ departs from the predicted curves for $f \gtrsim 10 \, \rm kHz$, except in the absence of electric noise. Our formulas were derived assuming a white spectrum for the random electric force. In the experimental setup, the spectrum of the electric signal is indeed flat but only up to some cutoff frequency where it decays to zero very fast. The computation of $\langle \overline{v}^2_f\rangle$ can be modified to take into account this cutoff frequency, at least numerically, as shown in the Supplementary Material\cite{supplemental}. Figure~\ref{fig:Tv_vs_tau} also shows the value of such calculation, using a cutoff frequency of $f_c=3\, \rm kHz$, with excellent agreement between theory and experiment. 

The measured cutoff frequency of the amplifier, and hence of the generated electric noise, is $f_{c,\text{amplifier}} = 10 \, \rm kHz$, in agreement with the specifications of our device, but significantly larger than the cutoff obtained from the fit of the data in Fig.~\ref{fig:Tv_vs_tau}. In order to elucidate the origin of this discrepancy, we measured the dynamic electrophoretic velocity $v_e^*(f_{\rm ac})$ in an additional experiment, from the analysis of the forced oscillations of the trapped bead as a function of the frequency $f_{\rm ac}$ of a sinusoidal electric field~\cite{Delgado2005Measurement,Pesce2013Optical}. In the inset of Fig.~\ref{fig:Tv_vs_tau}, we show that the electrophoretic response is almost flat up to the kHz region, where a strong decay above $f_{\rm ac}\simeq 3\, \rm kHz$ is clearly seen. This decay at frequencies below that of the cutoff frequency of the amplifier is probably due to the well-known alpha or concentration-polarization process, which predicts a relaxation of the mobility with a characteristic frequency $f_{\alpha}\simeq D^2/4\pi R^2$, $D$ being the diffusion coefficient of the counterions in the electric double layer and $R$ the radius of the bead. Under our experimental conditions, with no added salts in solution, the counterions are protons, and the expected characteristic frequency is $f_{\alpha}\simeq 3\, \rm kHz$~\cite{Delgado2005Measurement,Carrique2013Effects}. This value is in perfect agreement with the cutoff frequency obtained from the measurements of the TAV and the observed decay in the electrophoretic response. We note that the kHz region is inaccessible for the electroacoustic techniques typically used to quantify the dynamic electrophoretic velocity~\cite{rica2012electrokinetics}, and therefore the alpha relaxation of the mobility has only been marginally reported~\cite{Heinengen2010Dynamics}. 

One conclusion can be drawn. Our results demonstrate that, at sufficiently low sampling rates, we cannot distinguish, neither in the position nor in the velocity of the bead, between the effect of a random force or an actual thermal bath at higher temperature. From this result, we claim that our setup can be used as a simulator of thermodynamic processes at very high temperatures, as we show next. 


We now  investigate if it is possible to ascertain the average kinetic energy change of a microscopic system in thermodynamic quasistatic processes. We are interested in measuring the kinetic energy change along a process, averaged over many realizations, $\langle \Delta E_{\rm kin} \rangle = \frac{1}{2}m\Delta\langle v^2(t)\rangle$. To do so, we first compute the value of the TAV along a trajectory $x_t$ sampled at low frequency, $f=1\,\rm kHz$, and then estimate the mean square instantaneous velocity using 
\begin{equation}
\langle v^2(t)\rangle=L(f)^{-1} \langle \overline{v}^2_f(t)\rangle.
\label{eq:velBDC}
\end{equation}
This relation does not allow us to compute the instantaneous velocity from the TAV for a particular trajectory. It nevertheless gives enough information to evaluate changes in the average kinetic energy in any process where an external control parameter is modified, since the average kinetic energy at any time $t$ in a quasistatic process can be estimated as 
\begin{equation}
	\langle\Delta E_{\rm kin} (t)\rangle = \frac{1}{2} m \Delta\left(\frac{\langle \overline{v}^2_f(t)\rangle}{ L(f)}\right).
	\label{eq:EkinTAV}
\end{equation}

\begin{figure}
\includegraphics[height=4cm]{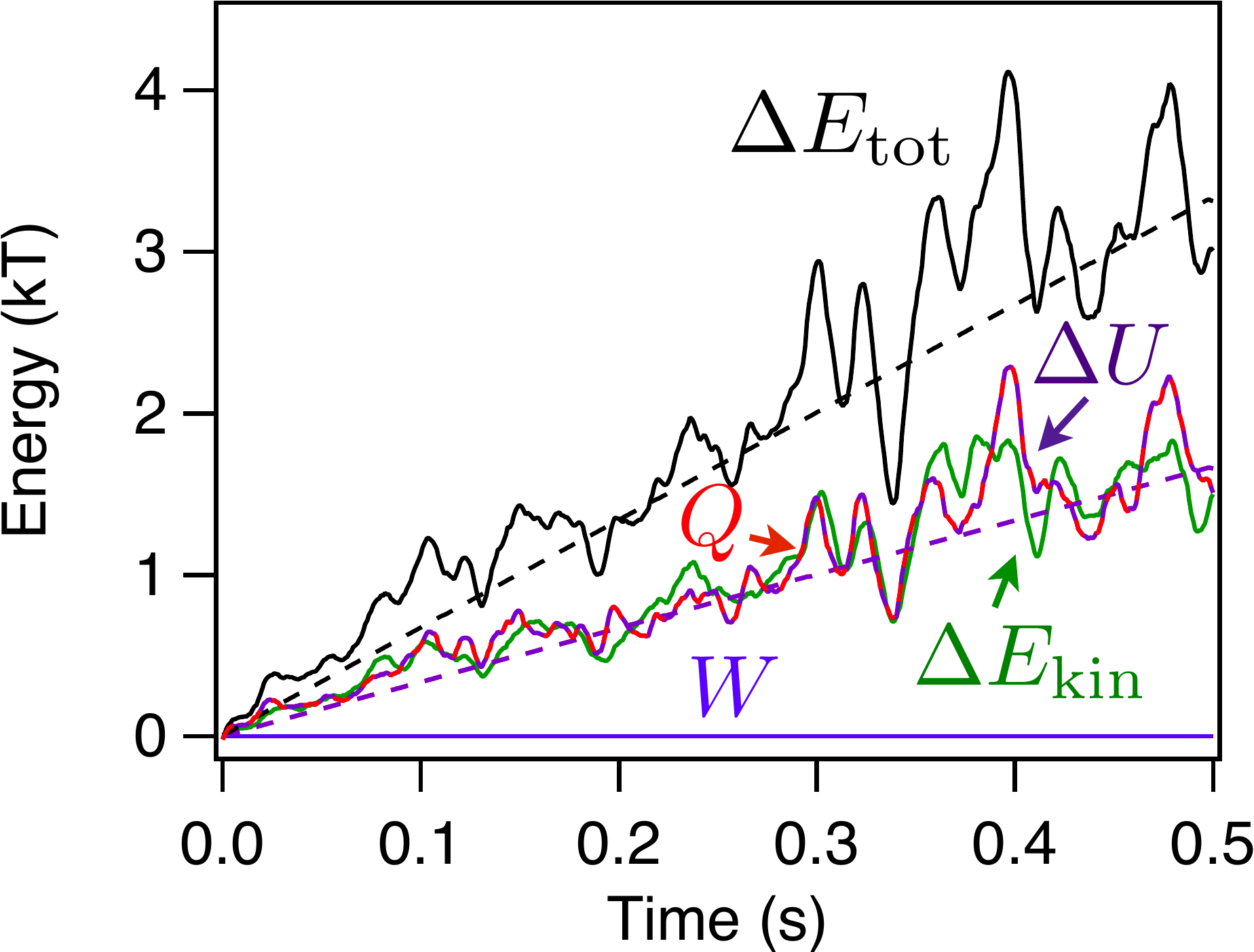} 
\caption{Experimental ensemble averages of the cumulative sums of thermodynamic quantities as functions of time in a non-isothermal process, where $T_{\rm kin}$ changes linearly with time from $300~\rm K$ to $1300~\rm K$ at constant $\kappa=(18.0 \pm 0.2)\,\rm pN/\mu m$:  Work (blue solid line), heat (red solid line), potential energy (magenta solid line), kinetic energy (green solid line) and total energy (black solid line).  Potential energy and heat are overlapped in the figure. Theoretical predictions are shown in dashed lines. All the thermodynamic quantities are measured from low-frequency sampled trajectories of the position of the particle, with $f=1\,\rm kHz$. Other parameters: $\tau=0.5\,\rm s$; ensembles obtained over $900$ repetitions.}
\label{fig:IC_energetics}
\end{figure}

We first implement a \emph{non-isothermal} process in which the stiffness of the trap is held fixed and the kinetic temperature of the particle is changed linearly with time. Figure~\ref{fig:IC_energetics} shows the experimental and theoretical values of the cumulative sums of the ensemble averages of the thermodynamic quantities. Heat and work are obtained from $1-\rm kHz$ measurements and $\Delta U = Q+W$. Average kinetic energy is extrapolated to high frequencies using Eq.~\eqref{eq:EkinTAV} and total energy is obtained as $\Delta E_{\rm tot} = \Delta U + \Delta E_{\rm kin}$. Since the control parameter does not change, there is no work done on the particle along the process. 
 The potential energy change satisfies equipartition theorem along the process, and heat and potential energy changes coincide, $\langle \Delta U(t)\rangle = \langle Q(t) \rangle = \frac{k}{2}[T_{\rm kin}(t)-T_{\rm kin}(0)]$. Our measurement of kinetic energy is in accordance with equipartition theorem as well, yielding $\langle \Delta E_{\rm kin} (t) \rangle = \frac{k}{2}[T_{\rm kin}(t)-T_{\rm kin}(0)]$. Adding the kinetic and internal energies, we recover the expected value of the total energy change of the particle, $\langle \Delta E_{\rm tot} (t) \rangle = k [T_{\rm kin}(t)-T_{\rm kin}(0)]$. 

\begin{figure}
\includegraphics[height=4cm]{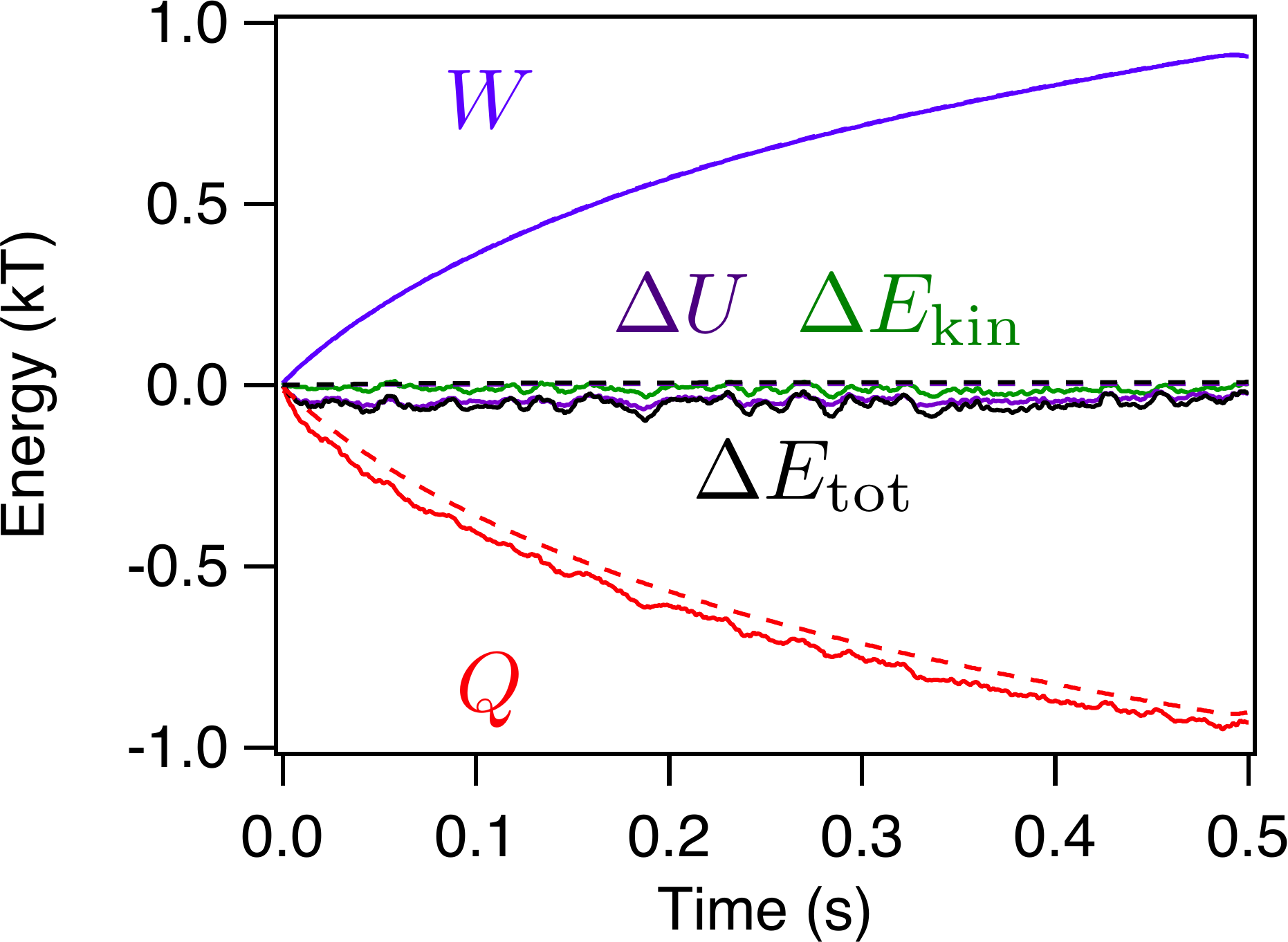} 
\caption{The same as Fig.~\ref{fig:IC_energetics}, but for the isothermal process, where $\kappa$ changes linearly with time from $(5.0\pm 0.2)\,\rm pN/\mu m$ to $(32.0\pm 0.2) \,\rm pN/\mu m$ at constant $T_{\rm kin}=T$.}
\label{fig:IT_energetics}
\end{figure}

In a second application, we realize an \emph{isothermal} process, where the external noise is switched off $(T_{\rm kin}=T)$ and the stiffness of the trap is increased linearly with time. Figure~\ref{fig:IT_energetics} shows the ensemble averages of the cumulative sums of the most relevant thermodynamic quantities concerning the energetics of the particle. As clearly seen, the experimental values of heat and  work coincide with the theoretical prediction for the case of isothermal quasistatic processes, i.e., $\langle W(t) \rangle = - \langle Q(t) \rangle = \frac{k}{2}T_{\rm kin}\ln\left(\kappa(t)/\kappa(0)\right)$. The potential energy change, $\langle \Delta U (t)\rangle = \langle W(t)\rangle + \langle Q(t)\rangle$ vanishes as expected for the isothermal case. We measure the kinetic energy change from the TAV using Eq.~\eqref{eq:EkinTAV}. Our estimation of kinetic energy change vanishes, in accordance with equipartition theorem $\langle \Delta E_{\rm kin} (t)\rangle = \frac{k}{2} \Delta T_{\rm kin} = 0$. The variation of the total energy, $\langle \Delta E_{\rm tot} (t)\rangle = \langle \Delta U (t)\rangle  +  \langle \Delta E_{\rm kin} (t)\rangle$ vanishes as well. We remark that despite the TAV is obtained from position data, it captures the distinct behavior of the velocity with respect to the position in the isothermal case, where $\langle x^2 (t) \rangle = kT_{\rm kin} / \kappa(t)$ changes with time whereas $\langle v^2 (t)\rangle = kT_{\rm kin}/m$ does not.

Figures~\ref{fig:IC_energetics} and~\ref{fig:IT_energetics} show the experimental values of kinetic energy changes of a microscopic system in quasistatic thermodynamic processes. Using Eq.~\eqref{eq:EkinTAV} we are able to access the complete thermodynamics of a Brownian particle in quasistatic processes. In non-isothermal processes, the First Law of thermodynamics reads $\Delta E_{\rm tot} = Q_{\rm tot}+W$, where $Q_{\rm tot} = Q + Q_v$ and $Q_v=\Delta E_{\rm kin}$ is the heat transferred to the momentum degree of freedom. A complete description of the Second Law and the efficiency of microscopic heat engines would also benefit from the measurement of $\Delta E_{\rm kin}$~\cite{Sekimoto2010}. In particular, a correct energetic characterization of artificial Carnot engines~\cite{Schmiedl2008,Bo2013} or Brownian motors~\cite{Reimann2002} would benefit from the measurement of $Q_v$.

To summarize, we have estimated the kinetic energy change of a single microscopic colloid with high accuracy, both in equilibrium condition as well as in both isothermal and non-isothermal quasistatic processes. This has been achieved from the measurement of the mean square instantaneous velocity of a Brownian particle inferred from trajectories of its position sampled at low frequencies. As a by-product of the measurement of the variance of the time averaged velocity, we have been able to quantify properties of the electrophoretic mobility of the particle in water such as the alpha relaxation frequency, which is typically unattainable from standard characterization techniques. This mobility relaxation limited the bandwidth of the applied white noise. In order to get a wider bandwidth, the precise knowledge of the mobility spectrum could be used to apply a non-white noisy signal, in such a way that the effective electric force on the particle is in turn white. Our tool could be extended to measure temperature gradients in inhomogeneous media and to evaluate the complete thermodynamics of nonequilibrium non-isothermal processes affecting any mesoscopic particle immersed in a thermal environment and trapped with a quadratic potential. For instance, one could measure the average heat transferred to the velocity degree of freedom   for Brownian motors described by a linear Langevin equation, such as molecular biological motors~\cite{lacoste2009fluctuation} and artificial nano heat engines~\cite{Blickle2011}.

ER, IM and RR acknowledge financial support from the Fundaci\'o Privada Cellex
Barcelona, Generalitat de Catalunya grant 2009-SGR-159, and from the MICINN (grant FIS2011-24409).
LD and ER acknowledge financial support from the Spanish Government (ENFASIS). LD acknowledges financial support from Comunidad de Madrid (MODELICO). We thank Antonio Ortiz-Ambriz, Juan M. R. Parrondo and F\'elix Carrique for fruitful  discussions. 


%

\end{document}